University of Canberra
Faculty of Information Sciences and Engineering

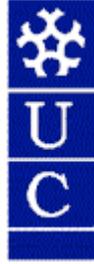

# Knowledge Management Systems Requirements Specifications

Omar Sultan Al-Kadi

*May 2003*

# Table of Contents

# Table of Contents (continued)

# Letter of Transmittal

The proposed report will discuss how to specify Knowledge Management System (KMS) requirements for report 2[1]. Where report 2 was on '*The design of a knowledge management architecture for the case study of the human resources of the commonwealth government agency'*.

This report was prepared from various information gathered form the World Wide Web, books, lectures and from my past knowledge and my own perspective to the discussed issue.

Report three is divided into several sections. Each of these sections is organised as follows:

- An introduction and a background section.

- An analysis that provides an in-depth look at of what possible problems that could be encountered. Through the gathering, evaluation and prioritisation of the specified KMS requirements for the agency.

- A cost / benefit analysis of the specified requirements of the KMSs in a way to insure that the proposed requirements would meet the goals which they should serve.

- Section 5 and 6 of the report will explain in detail and try to focus on the performance and the intended functionality of the KMS requirements.

- Measurement of success is a major goal that this report will try to present. Since what benefit could we get if this report did not present any means or procedures to display (sense and imagine) the expected outcome of these requirements.

- The last two sections will discuss how to improve the KMSs further (future trends) and a conclusion of what the report has presented.

This report is written to be comprehensive, through examining all of the areas that are important to address. The analysis and the focus on the problems are intended to be a stepping stone to assist the commonwealth government agency senior manager. This is done by suggesting some KMS requirements to be applied to the KM components and their architecture which was discussed in the previous report. The aim is to giving him a guide on how to implement these KMS requirements to achieve the required objectives and improve the agency's performance.

---

[1] *A brief summary of report 2 will be provided in section 2 in this report.*

## Abstract

*In recent years, Knowledge Management Systems (KMS) have drawn remarkable attention. However, there is no common understanding of how a knowledge management system should look like or where the corresponding research should be directed at. Based on a number of essential requirements that a KMS should satisfy, this report introduces some possible requirements for the commonwealth's KMS components forming the KMS architecture. Also, these requirements will be analysed through evaluating and measuring there functionality to produce a tangible outcome.*

## 1. Introduction

A *knowledge management system* (KMS) is the software framework (toolbox) that is intended to assist, via knowledge processing functions, those who desire to formulate and retrieve knowledge for different applications, such as system design and specification, term bank construction, documentation or ontology design for (multilingual) language processing (*Eagles, 1995*). The various tools of such a framework should help users to originate and organise ideas or understand and communicate ideas more easily and accurately than can be done with most current tools. A KMS is an integrated multifunctional system that can support all main knowledge management and knowledge processing activities, such as:

- Capturing;
- Organising;
- Classifying and understanding;
- Debugging and editing;
- Finding and retrieving;
- Disseminating, transferring and sharing knowledge (*Eagles, 1995*).

This report attempts to represent and focus on some of the requirements that KMSs need for our case. The aiming is to avoid some of the problems and bottlenecks that would encounter the users of these systems.

After having demonstrated our problems and the required systems to resolve them, we will move to the next stage of defining our KMS requirements through gathering, evaluating and prioritising them. Next a cost/benefit analysis of these requirements will be discussed.

The following two sections will discuss the expected performance and functionality of the proposed requirements, followed by a section that will try to measure the expected outcome in a sensible form.

Finally, this report will talk about some possible future trends, as how we can further develop these KMSs. This section will be followed by a conclusion.

## 2. Background

Initially, KMSs were addressed to solve some of the problems that were determined in report 1. One main task of the KMSs implemented is to search for specific information and try to employ it in a way that serves best the agency's goals and objectives (i.e. squeezing-out the knowledge from the available information). This is mainly done by the KMS components of the proposed architecture discussed in report 2. These systems could be hard knowledge management systems, as Information Retrieval Systems (IRS), Online Analytical Processing Systems (OLAPS), Electronic Records Managing Systems (ERMS), Artificial Intelligence (AI) & Expert Systems. Or these systems could be in a form of soft knowledge management system as Communities of practice (CoP). The next section will show how we can derive these requirements.

## 3. Deriving the KMS requirements

### 3.1 KMS requirements gathering

In order to gather the KMS requirements, we need to know accurately how a KMS should be. Hence, we can figure out what possible specification does each KMS miss, or how we can enhance the system by proposing new functions to it.

One of the basic appraoches to determine which requirements fit to which KMS is data colleciton. The goal of the data collection is to produce a formal document of KMS requirements through receiving feedback from the user of the KMS. This document of requirements must include the data needed by users as well as the use made of the data. This could be done through face-to-face interviews, on/off line survays, etc (see figure 1).Evaluators then should interview the users and then obtain even more detailed information from other sources, such as:

- Documents in the commonwealth agency that needs to use the product;
- Transactions that will be manageable by the product in the agency;
- Functional descriptions of business functions;
- Business rules;
- Scenario analyses.

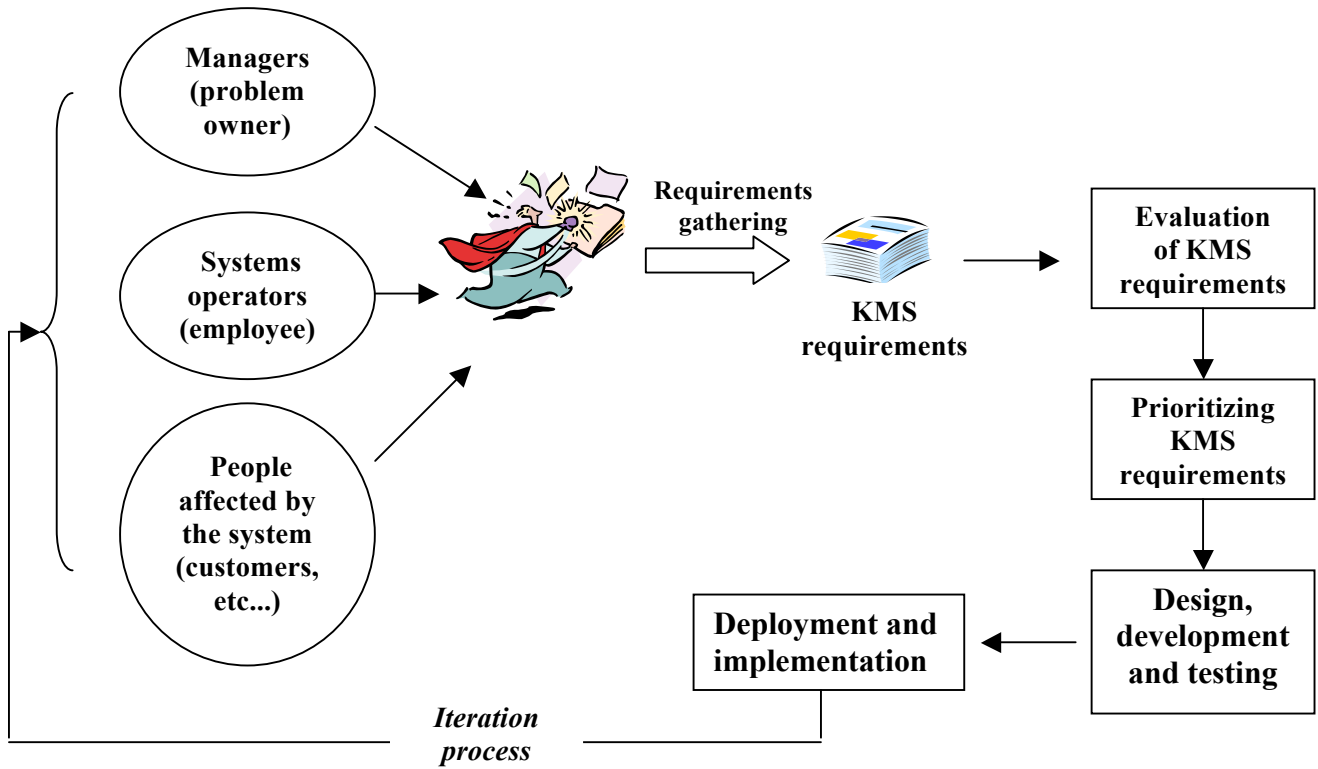

**Figure 1 process of deriving KMS requirements**

## 3.2 Evaluation of the selected requirements

The evaluation stages must follow the natural design process of a product to be evaluated, which begins by collecting and analysing the data regarding the application that aims at employing the product (purpose of use). The evaluation specification is then prepared and the technical evaluation follows (*Tansley & Hayball,1993*).

KMS requirenemnts can be evaluated (in a broad sense) according to the degree to which it satisfies certain conditions. In other words, when these requirements are introduced, do they improve the KMS by:

- Ease of use;
- Domain independence;
- Efficient and effective indexing support (machine aided or automatic);
- Response efficiency;
- Presentation of results;
- User and database (knowledge base) services.

There is one point we should emphasise on, which is, although the commonwealth agency's employees provide valiable feedback in the process of KMS requirements gathering. However, user interviews can only help to a limited extent for improving knowledge management systems and information retrieval procedures. This is because users can only evaluate and require what they know and what they can imagine; new application technologies have to be initiated or introduced first .

### 3.3 Prioritisation of the selected requirements

After the process of KMS requirements gathering and evaluating is accomplished comes prioritisation. It is crucial for organizations to determine which KMS requirements for which KMS needs to be implemented first. Some KMSs are considered vital and are an important asset to keep the agency running could have the highest priority; while other systems can wait (have a lower priority).

For example, in the case of the commonwealth agency, there are some KMS which are considered important to its human resources department. A result of their absence could case the department to stop running effectively; hence KM requirements of these systems need to be driven first.

KMSs as IRS, EDMS and OLAP have the highest priority among the rest of the KMS components in the KMS architecture. While the other three proposed systems in the KM architecture (data mining, expert systems and CoP) could be rated at a lower priority level.

## 4. A cost and benefit analysis

Completing the prioritizing process, the cost and benefit analysis quantifies financially the actions proposed for each application. To be useful for KMS prioritizing, the analysis must include the cost of failing to do the work and the benefits of intangible results (*Yourdon, 1989*). To reach final conclusion in most cases, highly refined data is not necessary. Most analysis can proceed if the cost and benefits are predictable within 20 percent or so. Additional refinement may be required to resolve close decision. This analysis must incorporate the time value of money to ensure financially integrity (*Rackoff, 1985, p.285*).

Bellow in table 1 is a cost and benefit analysis for the proposed KMS requirements. This table is composed of seven attributes for each KMS to be compared to. Which are:

a) *User and It satisfaction*: as previously described in figure 1, to be determined through face-to-face interviews, on/off line surveys, etc…
b) *Strategic and Tactical importance*: to be determined through the agencies senior managers.
c) *Time required for development*: the KMS evaluators in conjunction with other IT people who are appointed to develop such systems such estimate the time required for determination and implementation of the KMS requirements.
d) *Cost and Benefits*: to be specified according to the KMS tangible and intangible implementation and outcome.

In table 1, some estimated ratings are given to the various attributes; for which **6** stands for very high priority (highest priority) while **1** is for the lowest priority. Similarly for the time required for development of these KMS requirements; **6x** needs the longest time for development, while **x** is the shortest to develop. The same applies to cost and benefits.

**Table 1 Cost/benefit analysis for the KMS requirements**

| KMS name | User Satisfaction | IT Satisfaction | Strategic importance | Tactical importance | Time required | Cost | Benefits |
|----------|-------------------|-----------------|----------------------|---------------------|---------------|------|----------|
| IRS | | | 6 | 6 | $3x$ | 80K | 180K |
| OLAP | | | 3 | 4 | $4x$ | 100K | 130K |
| EDMS | | | 2 | 5 | $2x$ | 50K | 70K |
| Data Mining | | | 4 | 2 | $5x$ | 50K | 100K |
| Expert System | | | 5 | 1 | $6x$ | 100K | 200K |
| CoP | | | 1 | 3 | $x$ | 10K | 30K |

**To be determined by the agency's employees (managers, HRM staff, technical 'IT' staff, etc…) and by customers.**

Other attributes can be added as well to this table in order to refine the data in the cost/benefit analysis for the purpose of decision improvement. An example of other attributed could be performance, availability, security, and modifiability.

# 5. Features of KMS requirements

In general, a KMS should serve everybody who is involved in processes of understanding, evaluating and organizing the business. Due to the nature of these tasks, the primary focus groups include consultants (internal and external), new employees who have to understand the company in general and their task in particular, executives, system analysts as well as customers and suppliers that participate in cross-organizational business processes. A KMS should provide these groups with relevant knowledge. With respect to the notion of knowledge introduced before we can now describe more specific requirements, a KMS should fulfill.

The following are some of the features of a KMS requirements mentioned in a paper presented by Professor Ulrich Frank (*Frank, 2001*) which are applicable to our case.

### 5.1 Emphasis on Concepts and Reason

A KMS should offer definitions of concepts that are needed for the description and analysis of a corporation. To give a few examples for such concepts: corporate strategy, organizational unit, business process, task, employee etc. Note that these concepts are usually not defined independently from one another. In contrast to a traditional information system, a KMS should allow to answer questions that refer to concepts, for instance:
• What is a business process?
• What are the concepts to describe a new type of business process?
• What are the organizational implications of a strategy that is aimed at cost leadership?

## 5.2 Re-use of Existing Knowledge

Although there is no unified terminology for the description of corporate knowledge, there are a number of elaborated and well-documented concepts available or provided, for instance, by textbooks. This is also the case for the documentation of relevant causal relationships. A knowledge management system should provide an adequate body of existing knowledge. This is for various reasons. The re-use of knowledge does not only contribute to the economics of a KMS. It should also improve the overall quality of its content. In addition to that it fosters communication by referring to a body of knowledge many people are familiar with.

## 5.3 Support of Multiple Perspectives

In order to support different users and different tasks, a KMS should provide various perspectives on the knowledge it stores. Managing complexity recommends offering different levels of detail. For instance, sometimes it will be sufficient to get a description of a business process that is restricted to an outline of the temporal relationships between high-level tasks. In other cases it may be important to provide a comprehensive description of every task within the process as well as of the required resources.

## 5.4 Integration with Information

With a KMS there is emphasis on storing knowledge rather than information. However, information cannot be completely neglected. This is for two reasons. Firstly, the distinction between knowledge and information depends in part on subjective judgment. For this reason the complete exclusion of information is not possible. Secondly: While it is obvious that knowledge adds value to related information, it is also the case that information adds value to knowledge. Since knowledge is rather abstract, having access to corresponding instances will help many people to develop a proper understanding. Therefore a KMS should support the integration of knowledge with information.

## 5.5 Support of Awareness

In order to foster organizational learning, a KMS should support the dissemination of knowledge. Whenever its content gets updated, users that are interested in the corresponding topics should be notified. For this purpose a user should be able to subscribe to certain types of knowledge or - more general - of content.

# 6. Tasks and Functionality of the KMS requirements

This section of the report focuses on how the selection of KMS requirements would serve and improve the KMSs performance. But before going into details for each system, first we need to know what sorts of requirements can make a KMS effective and efficient.

Effective knowledge management systems must be:

- **Scalable**:  must be able to support a large number of users and a robust, industrial strength database;
- **Extensible**: capable of expanding as needed by the organization;
- **Compliant with industry standards**: allowing companies to leverage existing resources;
- **Secure**;
- **Relevant and Timely**;
- **Collaborative**: although many efforts start with a single department of group, the best KMS grow to encompass input from across the organization;
- **Powerful off-line analysis**;
- **Allow for complex queries**;
- **Fast and easy to administer and deploy**;
- **Flexible**: The technology should be able to handle knowledge of any form, including different subjects, structures and media. It should be able to handle forms which do not as yet have been defined;
- **Heuristic**: The systems should learn about both its users and the knowledge it possesses as it is used. Over time, its ability to provide users with knowledge should improve. For example, if the solution deals with many requests on a particular subject, it should learn how to assist users in more depth on that subject;
- **Suggestive**: The solution should be able to deduce what a user's knowledge needs are and suggest knowledge associations that he is not able to do himself (*Abeljaber, Ioannou, Maor, Razo, Tribolet, 1998*).

## 6.1 Information Retrieval Systems (IRS)

The major tasks and functions required from a IRS for our case is to:

- Reference or primary information (bibliographic information *vs.* complete text);
- Form of information (text *vs.* facts, data, etc.);
- Combination of different media (text *vs.* multimedia);
- Life-time (e-mail *vs.* documentation);
- Copyrights and licences (own information *vs.* foreign information);
- Encapsulation of information (unrestricted access *vs.* controlled access, e.g. CD);
- Text length (short texts *vs.* long texts, e.g. product manuals);
- Text form (text for print media *vs.* text for electronic media);
- Linearity (linear text *vs.* hypertext).

There are several approaches to improve IR systems in general, particularly approaches oriented towards relevance analysis. Among these are, for example, dialogue components (menu-based or natural language based) and statistical approaches. The former are mainly used to improve the use of a given query language according to pre-defined user profiles (novice, expert, etc.), whereas the latter will improve the relevance analysis directly (*Zhang & Mostafa, 2002*).

## 6.2 Online Analytical Processing Systems (OLAP)

A typical OLTP system is characterized by having large numbers of concurrent users actively adding and modifying data. The database represents the state of a particular business function at a specific point in time. However, the large volume of data maintained in many OLTP systems can overwhelm an organization. As databases grow larger with more complex data, response time can deteriorate quickly due to competition for available resources. A typical OLTP system has many users adding new data to the database while fewer users generate reports from the database. As the volume of data increases, reports take longer to generate (*Microsoft, 1998*).

Therefore there are some requirements that this system should undertake. The OLAP system should have the following requirements in order to resolve some of the problems that might emerge, which are:

- *Multiple data sources*: An intuitive multidimensional data model that makes it easy to select, navigate, and explore the data.
- *Time oriented*: providing a very fast response to users queries.
- *Ad hoc queries*: A powerful tool for creating new views of data based upon a rich array of ad hoc calculation functions.
- *Aggregated and summarized*: Pre-aggregation of frequently queried data, enabling a very fast response time to ad hoc queries.
- *Security*: Technology to manage security, client/server query management and data caching, and facilities to optimize system performance based upon user needs.

## 6.3 Electronic Record Managing Systems (ERMS)

There are a number of requirements that the ERMS should specify. Bellow is an explanation of some of these ERMS requirements which are suitable to our case.
### 6.3.1 Classification Scheme
A classification scheme lies at the heart of any ERMS. It defines the way in which the electronic records will be organized into electronic files, and the relationships between the files.

  i.   The ERMS must support metadata for files and classes in the classification scheme; and after a record has been captured the ERMS must restrict the ability to add to or amend its metadata to Administrators. Also the ERMS must record the date of opening of a new class or file within the file's metadata.

ii.   The ERMS should support an optional class and file naming mechanism which includes names (e.g. persons' names) and/or dates (e.g. dates of birth) as file names, including validation of the names against a list (*Cornwell, 2001*).

## 6.3.2 Controls and Security

The ERMS must allow the Administrator to limit access to records, files and metadata to specified users or user groups. Additionally, ERMS must allow the Administrator to attach to the user profile attributes which determine the features, metadata fields, records or files to which the user has access.  The attributes of the profile will:

- prohibit access to the ERMS without an accepted authentication mechanism attributed to the user profile;
- restrict user access to specific files or records;
- restrict user access to specific classes of the classification scheme;
- restrict user access according to the user's security clearance;
- restrict users access to particular features (e.g. read, up-date and/or delete specific metadata fields);
- deny access after a specified date.

The ERMS must keep an unalterable audit trail capable of automatically capturing and storing information about:

- all the actions that are taken upon an electronic record, electronic file or classification scheme;

- the user initiating and or carrying out the action;

- the date and time of the event (*Cornwell, 2001*).

a) *Backup and Recovery*
The ERMS must provide automated backup and recovery procedures that allow for regular backup of all or selected classes, files, records, metadata and administrative attributes of the ERMS repository.

b) *Tracking Record Movements*
A tracking feature is needed to record the change of location for both ease of access and to meet regulatory requirements.

c) *Authenticity*
The ERMS must restrict access to system functions according to user's role and strict system administration controls. Additionally, where possible and appropriate, the ERMS should be able to provide a warning if an attempt is made to capture a record which is incomplete or inconsistent in a way which will compromise its future apparent authenticity.

The third authentication requirements is that ERMS must prevent any change to the content of the electronic record by users and Administrators (except where change is part of the business and/or documentary process, as discussed elsewhere in this specification).

*d) Security Categories*

The ERMS must allow security categories to be assigned to records. Also ERMS must allow, but not necessarily require, security categories to be made up of one or more subcategories, such as:

   i.    *Class* (Top Secret, Secret, Confidential, Restricted, Unclassified)
   ii.   *Caveat* (CEO Eyes Only, CIO Eyes Only, etc…)
   iii.  *Descriptor* (Commercial, Personnel, Management, Audit and Accounts)

**6.3.3 Retention and Disposal**

The ERMS must provide a function that specifies retention schedules, automates reporting and destruction actions, and provides integrated facilities for exporting records and metadata.

**6.3.4 Capturing Records**

The ERMS record capture process must provide the controls and functionality to:

- register and manage all electronic records regardless of the method of encoding or other technological characteristics;

- ensure that the records are associated with a classification scheme and associated with one or more files;

- integrate with application software that generates the records;

- validate and control the entry of metadata into the ERMS (*Cornwell, 2001*).

**6.4 Data Mining Systems (DMS)**

DMS are techniques which search for relationships, patterns, and trends which, prior to the search were not known to exist or were not visible. Some DMS requirements specifications are:

*a) Associations*

Identifying affinities that exist among the collection of items in a given set of records. For example, 80% of all records that contain A, B and C also contain D and E.

*b) Sequential Patterns*

Identify frequently occurring sequences from given records

*c) Classifying*

❑      Identify *a priori* certain mutually exclusive classes.
❑      Identify a set of attributes that discriminate among the classes.
❑      Differentiating between frequent, moderate and infrequent occurrences.

*d) Clustering*
- ❑       Identifying categories
- ❑       Natural grouping of customers by processing all the available data about them (*Rai & Storey, 2001*).

## 6.5 Communities of practice (CoP)

There are two main requirements for CoP to operate effectively, which are:

- People should have a strong sense of identity tied to the community (e.g., as technicians, salespeople, researchers and so on).
- As the practice itself is not fully captured in formal procedures, people should learn how to do what they do and become seen as competent in the course of doing it in concert with others.

Additional requirements are:

- Continuing mutual relationships, harmonious or conflicting (i.e., regular, work-related interactions, rough or smooth)
- Shared ways of doing things together (i.e., common practices and beliefs about best practices)
- A rapid flow of information between and among members
- Quick diffusion of innovation among members (e.g., rapid transfer of best practices)
- Conversations need to come quickly to the point (i.e., no lengthy lead ins)
- Problems should be quickly framed (i.e., a common understanding of the milieu in which they all operate)
- A widespread and shared awareness of each others' competencies, strengths, shortcomings and contributions
- An ability, concentrated or distributed, to assess the effectiveness of actions taken and the utility of products produced
- Common tools, methods, techniques and artifacts such as forms, job aids, etc.
- Common stories, legends, lore, "inside" jokes, etc.
- A shared, evolving language (e.g., special terms, jargon, "shortcuts" such as acronyms, etc.)
- Behavior patterns that signify membership (e.g., gestures, postures, and even seating patterns in the cafeteria)
- Perspectives reflected in language that suggest a common way of viewing the world, such as shared analogies, examples, explanations, etc. (*Nickols, 2000*)

# 7. Measurement of success

Measurement refers to all KM activities that measure, map and quantify corporate knowledge and the performance of KMSs.

Measures that are essential for IRS, OLAPS and ERMS can be summarised as:

- Investigation amount (time and costs);
- Procurement amount (time and costs);
- Evaluation amount (time, redundancy of information).

Other measurements procedures for KMS requiremnts to insure their effectiveness can be identified as follows:

*a) Measurement of Precision*: A text presented as an answer to an inquiry should contain only relevant information.
*b) Measuremnt of Recall*: All texts containing relevant information should be found and presented as an answer to an inquiry.
*c) Measurement of Specificity*: Information requested (by the user) or offered by the texts stored in the database should be expressible without limitations. It should be possible to specify any subject and any detail of the requested information the system has to search for (*Lopez, 2001*).

## 8. Future trends

Organization seeks always to be up-to-date in various fields and KMSs are one of those fields. It is recommended that the commonwealth government agency to keep its KMS up-to-date. That does not mean buying anything and everything new emerges in the market and abandoning the old, but through keeping the KMS requirements iteration cycle running (see figure 1 in section 3.1).

There is also on-going work in the field of introducing natural language processing technology into IR OLAP systems, which obviously would be a further improvement of these systems (*Salton & Buckley & Smith, 1998*). In addition, future KMSs will have to take into account new styles of publishing and electronic communication, including capabilities investigated in the field of *virtual reality* (*Sherman & Judkins, 1993*).

## 9. Conclusion

Current knowledge management systems, in particular those in the field of information retrieval, are too narrow in many respects. For example, one application, one type of user, one type of knowledge representation, one type of knowledge operation, etc. Also they are too hard to use. For example, specialised knowledge is needed and long training curves are necessary and are not widely known or available. Therefore users need to provide feedback in order to improve the KMSs overall performance, which is done through specifying each KMS requirements.

This report specifies and derives some desirable KMS requirements for the human resources departments of the commonwealth government agency. Then provides techniques to sense its intangible benefits through various techniques, such as, measurement of success and cost/benefit analysis. For which, all of the above proposed KMS requirements should serve its main purpose, which is smoothining the agency's way of work and improving it performance effeciently and effectively.